\documentclass[lettersize,journal]{IEEEtran}
\usepackage{amsmath,amsfonts}
\usepackage{algorithmic}
\usepackage{algorithm}
\usepackage{array}
\usepackage[caption=false,font=normalsize,labelfont=sf,textfont=sf]{subfig}
\usepackage{textcomp}
\usepackage{stfloats}
\usepackage{url}
\usepackage{verbatim}
\usepackage{graphicx}
\usepackage{cite}
\usepackage{xcolor}
\usepackage{orcidlink}
\usepackage{pdfpages}
\begin{document}
\begin{titlepage}
\noindent {\Large \textbf{IEEE Copyright Notice}} \\

\noindent {\large © 2025 IEEE. Personal use of this material is permitted. Permission
from IEEE must be obtained for all other uses, in any current or future media, including reprinting/republishing this material for advertising or promotional purposes, creating new collective works, for resale or redistribution to servers or lists, or reuse of any copyrighted component of this work in other works.}
\end{titlepage}
\title{Active Eye Lens Dosimetry With Dosepix:\\Influence of Measurement Position and Lead Glass Shielding}

\author{Leonie Ullmann$^{\orcidlink{0009-0001-0521-0782}}$, Florian Beißer$^{\orcidlink{0009-0005-8068-3327}}$, Rolf Behrens$^{\orcidlink{0000-0002-4905-7791}}$, Stefan Funk$^{\orcidlink{0000-0002-2012-0080}}$, Gerhard Hilgers$^{\orcidlink{0009-0004-1400-7876}}$, Oliver Hupe$^{\orcidlink{0000-0001-6561-3375}}$, Jürgen Roth$^{\orcidlink{0000-0002-0679-699X}}$, Tom Tröltzsch$^{\orcidlink{0009-0003-9955-0976}}$, Hayo Zutz$^{\orcidlink{0000-0002-0907-6509}}$, and Thilo Michel$^{\orcidlink{0000-0002-2960-0611}}$ 
\thanks{This work did not involve human subjects or animals in its research.}
\thanks{Leonie Ullmann, Florian Beißer, Stefan Funk and Thilo Michel are with the Erlangen Centre for Astroparticle Physics, Friedrich-Alexander-Universität Erlangen-Nürnberg, 91058 Erlangen, Germany (e-mail: leonie.ullmann@fau.de).

Rolf Behrens, Gerhard Hilgers, Oliver Hupe, Jürgen Roth, and Hayo Zutz are with Physikalisch-Technische Bundesanstalt, 38116 Braunschweig, Germany.

Tom Tröltzsch was with the Erlangen Centre for Astroparticle Physics, Friedrich-Alexander-Universität Erlangen-Nürnberg, 91058 Erlangen. He is now with the Karlsruhe Institute of Technology (KIT), 76131 Karlsruhe, Germany.}}



\maketitle

\begin{abstract}
In this work, the effect of the measurement position on measurements of \textit{\textbf{H}}$_\text{\textbf{p}}$\textbf{(3)} of a new active eye lens dosemeter prototype based on the Dosepix detector is examined. A comparison between measuring directly in front of the eye and measuring at the side of the head of an Alderson phantom showed no significant influence on the resulting \textit{\textbf{H}}$_\text{\textbf{p}}$\textbf{(3)} for different radiation qualities and angles. In addition, to account for the absorption effect of radiation safety glasses, pieces of lead glass were attached to the front and side of the dosemeter.
The challenging geometry of the human face, when additionally equipped with absorbing glasses and irradiated at different angles of radiation incidence, led to a complex irradiation setup, whereby the dosemeter cannot be placed at the real point of interest, i.e. directly in front of the eye. Corresponding effects and consequences for radiation protection measurements have been investigated by using a human like Alderson head phantom as well as thermoluminescent dosemeters (TLDs) and the active eye lens dosemeter prototype based on the Dosepix detector. 
For specific angles, the radiation bypassed the radiation safety glasses and lead glass pieces, leading to an increase in the measured \textit{\textbf{H}}$_\text{\textbf{p}}$\textbf{(3)}. Compared to TLDs behind radiation safety glasses, measurements with the lead glass shielded prototype resulted in lower values for \textit{\textbf{H}}$_\text{\textbf{p}}$\textbf{(3)}. Furthermore, the results did not reproduce previous findings where larger dose values were found for Dosepix behind lead glass pieces than for the TLDs behind radiation safety glasses. A possible reason might be that the dimensions of the lead glass pieces are not representative of the radiation safety glasses in front of the eye but, ultimately, it is not yet clear what the main reason for the deviation is. Therefore, it is advisable to test the same methodology in future investigations with other eyewear models and lead glass pieces to investigate whether similar behaviors occur.
\end{abstract}


\section{Introduction}
\IEEEPARstart{T}{he} lens of the eye is one of the most radiation sensitive tissues in the human body. Excessive exposure to X-rays can cause severe damage to the eye lens, such as cataractogenesis \cite{ICRP118, Bitarafan, Bjelac, Elmaraezy, Vano}. To prevent such damage, there are strict limits the dose to the eye lens must not exceed \cite{ICRP118}. The recommendations for these limits have been adjusted in 2011 by the International Commission on Radiological Protection (ICRP) to an average dose of 20\,mSv/a within a 5-year period for occupational exposure, while the dose within a single year must not exceed 50\,mSv. In workplaces such as interventional medical procedures, the eyes are particularly exposed to radiation, and the dose to the lens of the eye is especially critical with regard to exceeding the dose limit. To ensure that these limits are not exceeded, a monitoring system is essential. 

The relevant quantity in the context of eye lens dosimetry is the operational quantity $H_\text{p}(3)$. The International Commission on Radiation Units (ICRU) defines it as the equivalent personal dose at a depth of 3\,mm in ICRU soft tissue \cite{ICRU1, ICRU2}. This quantity is used to estimate the dose deposited in the lens of the eye \cite{PTBDos23}, i.e. the equivalent dose in the lens of the eye.

A prototype of an active dosemeter for the eye lens was developed at the Erlangen Centre for Astroparticle Physics (ECAP) as part of Friedrich-Alexander-Universität Erlangen-Nürnberg (FAU) \cite{FloPaper}. This dosemeter is based on the hybrid pixel detector Dosepix and provides results within the IEC criteria in $H_\text{p}(3)$ measurements \cite{FloPaper}. However, it is unclear how the behavior of the dosemeter changes under more realistic conditions. So far, the eye lens dosemeter prototype has only been tested at the center front of a water cylinder phantom \cite{FloPaper} in accordance with corresponding type test standards by IEC such as IEC 61526:2024 \cite{IEC61526}. However, this does not represent the routine use since the dosemeter will be worn at the side of the head. This work examines the differences between those two measurement positions. Moreover, medical staff often use radiation safety glasses, which were not taken into account in previous investigations. In order to reproduce the shielding effect of such glasses, this work investigates the effectiveness of using lead glass pieces attached to the dosemeter.

\section{The Dosepix Detector}
The detector used for the active eye lens dosemeter is Dosepix \cite{Winnie} which is a hybrid energy-resolving and photon-counting pixelated X-ray detector with no dead time. Detailed information on the detector is provided in \cite{WinniePhD, Zang, Ritter}. Dosepix was developed in collaboration between Erlangen Centre for Astroparticle Physics (ECAP) as part of Friedrich-Alexander-Universität Erlangen-Nürnberg (FAU) and the European Organization for Nuclear Research (CERN). Due to its hybrid design, the radiation-sensitive layer is bump-bonded to the application-specific integrated circuit (ASIC). A p-in-n doped silicon sensor of 300\,µm thickness is used for the presented active eye lens dosemeter. Both ASIC and sensor are divided into $16 \times 16$ square pixels with a pixel pitch of 220\,µm. The pixels of the upper two and lower two rows of the sensor have an edge length of 55\,µm, resulting in non-sensitive areas between these pixels. The remaining $12 \times 16$ pixels of the sensor have an edge length of 220\,µm. This results in a total sensitive area of 9.49\,mm$^2$. In this work, only the results of the large pixels are considered, as these are sufficient in themselves to fulfill all the tested approval requirements \cite{FloPaper}.

\section{The Eye Lens dosemeter}
The prototype of the active eye lens dosemeter used for the measurements contains a Dosepix detector placed 2.5\,mm behind the front wall of the 3D-printed casing. To ensure a stable position and prevent possible damage to the sensor, a PMMA ring with a diameter of 18\,mm and a thickness of 2.5\,mm is fixed around the detector. The read-out hardware is connected to the Dosepix detector via a flat cable. The eye lens dosemeter is intended to be worn at the side of the head, so that the Dosepix detector faces to the front. Detailed information as well as the results of performance tests are provided in \cite{FloPaper}.

During the measurement process, each detector pixel compares incoming energy depositions to 16 freely adjustable energy bins and sorts them accordingly. The lowest bin edge is set to 12\,keV. The highest bin is used as an overflow bin and contains all events for which the deposited energy exceeds 150\,keV. The counts in each bin $i$ are summed up to the total number of entries $N_i$, which is then multiplied with a corresponding dose conversion factor $k_i$. The weighted sum over all 16 bins determines the equivalent dose of the eye lens $H_\text{p}(3)$. 
\begin{equation}
\label{eq:hp3}
H_\text{p}(3) = \sum^{16}_{i=1} k_i N_i
\end{equation}
The dose conversion factors $k_i$ used to obtain the dose from the detector signal were determined by a combination of measurements and radiation transport simulations and yield reliable results for photon radiation fields with energies up to about 250\,keV \cite{FloPaper}. The determination of these dose conversion factors represent the calibration of the device. The statistical uncertainty of the dose is estimated from Poisson statistics with the law of uncertainty propagation under the assumption of independent $N_i$ \cite{GUM, Dennis}. 
\begin{equation}
\label{eq:errhp3}
\sigma_{H_\text{p}(3)} = \sqrt{\sum^{16}_{i=1} k_i^2 N_i}
\end{equation}
In the following, only statistical uncertainties are considered. Other uncertainties, e.g. resulting from the positioning, have no significant influence on the results due to the detailed adjustment at the measurement facilities. \\
It should be noted that there is only a single set of dose conversion factors for all Dosepix detectors, which was derived from measurement data from a specific detector (Dpx 1 in the following chapters) \cite{FloPaper}. Since there are variations between individual detectors, this leads also to different results regarding reconstructed dose. However, these differences are within the limits of 0.71 and 1.67 stated in IEC 61526:2024 \cite{IEC61526}, as shown in \cite{FloPaper}. These limits are not symmetrical around the ideal response of 1.0, since an underestimation of the deposited dose has more severe consequences than an overestimation, which is taken into account for the definition of the response limits.

\section{Measurement position}
\subsection{Methods}
For official type approval, an eye lens dosemeter has to be placed centered in front of a water filled PMMA cylinder phantom with a diameter of 20\,cm \cite{phantom1, phantom2}. However, to mimic routine use situations, the following measuring setup was used.

\begin{figure}
\centering
\includegraphics[width=2.9in]{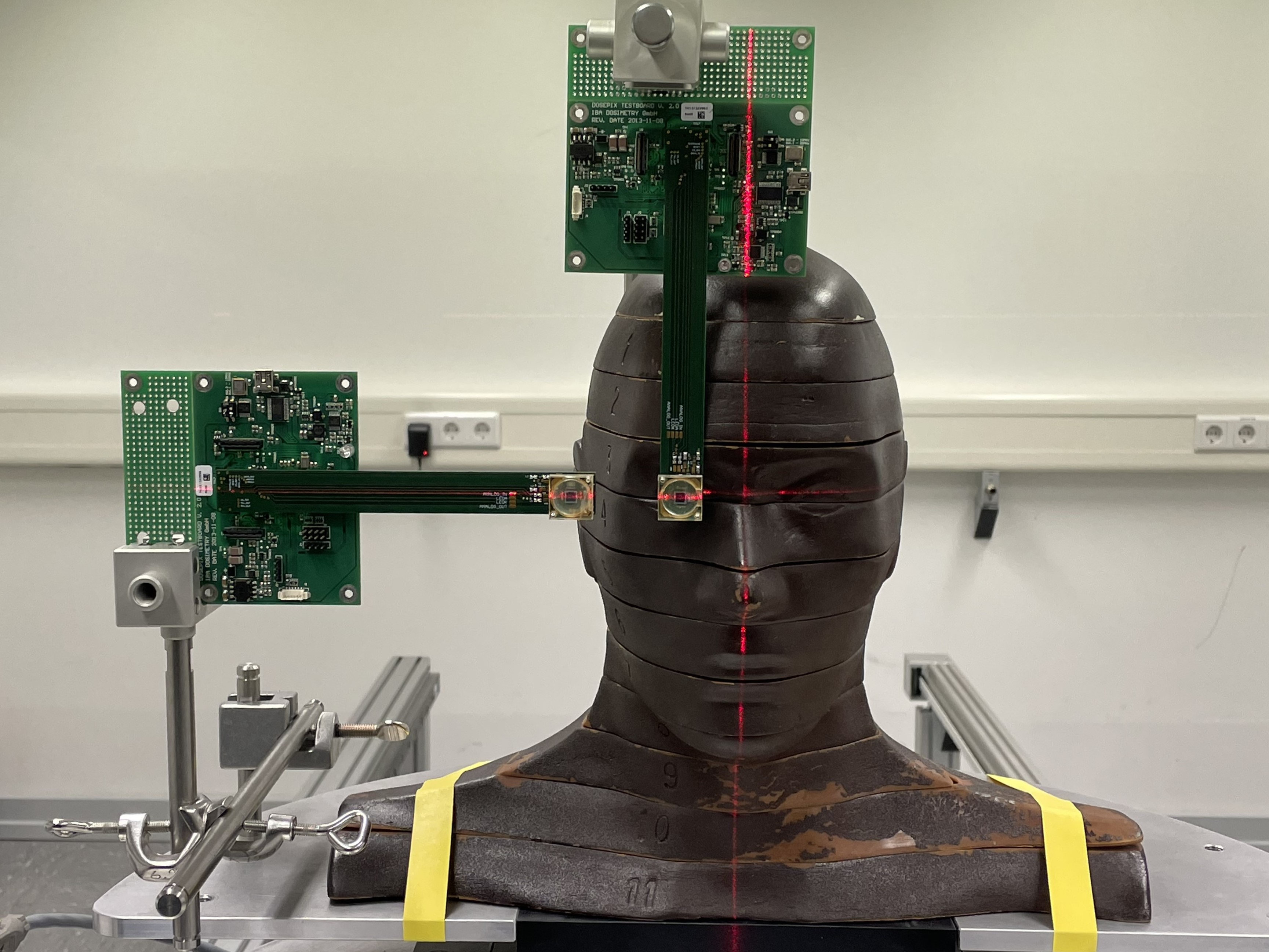}
\caption{Setup for comparative position measurements. Two Dosepix detectors are positioned on a connector extension, which is attached to the read-out hardware. One detector is placed on the eye of the phantom. The other detector is placed to resemble the position of the eye lens dosemeter prototype. Irradiation at $0^\circ$ means the radiation comes from the front onto the phantom.}
\label{fig:setup_kopf}
\end{figure}

Fig.~\ref{fig:setup_kopf} shows an image of the setup for the measurements at the facilities of Physikalisch-Technische Bundesanstalt (PTB). To compare the values of $H_\text{p}(3)$ measured on the eye and at the side of the head, two Dosepix detectors (numbered as Dpx 1 and Dpx 2) were placed on an Alderson phantom \cite{PhantomPaper}. This phantom allows for a more representative position of the dosemeter as it would be worn by the user. It should be noted that neither of the two detectors is installed in a dosemeter prototype, but is directly connected to the read-out hardware. To minimize the amount of material in front of the phantom, a connector extension is installed between Dosepix and its read-out hardware. Both detectors are in the same plane perpendicular to the beam axis. The reference point according to ISO 29661 \cite{ISO29661} is set to the nasal root of the Alderson phantom at 2.5\,m distance to the X-ray tube. The detector at the side is placed at a distance of 7.0\,cm to the reference point, the detector on the eye at a distance of 2.5\,cm. The detector on the eye was placed as close to the surface of the phantom as possible. Due to the hardware, a distance of 1.5\,cm was between sensor and surface.

The setup was irradiated with radiation qualities of the narrow-spectrum series (N-series) from N-20 to N-300 (with mean energies from 16\,keV to 248\,keV), with one irradiation per quality. Detailed information regarding the qualities' characteristics is given in ISO 4037-1 \cite{ISO40371}. This series of radiation qualities was chosen for comparability reasons with previous measurements \cite{FloPaper, Dennis}. The whole setup was illuminated homogeneously. To account for possible differences between the two detectors, the measurements were repeated where the detectors, including the respective hardware, were interchanged. Therefore, both detectors measured at both positions.
Additionally, the angle of irradiation was varied between 0$^\circ$ to $\pm 30^\circ$, $\pm 60^\circ$ and $\pm 75^\circ$ for the measurements with N-30, N-60, N-100 and N-200. Negative angles denote a clockwise rotation and positive angles a counterclockwise rotation. It should be noted that only the rotation regarding left/right anterior oblique view was examined since the setup does not enable a rotation in cranial/caudal direction. Since the latter case is common in the medical field, future studies should examine the behavior for angles in cranial/caudal directions. However, the influence is expected to be small since the angular dependency of the Dosepix detector is identical for horizontal and vertical rotation. The irradiated dose was set to 167\,µSv with a measurement time of 100\,s for all measurements. This reference dose was chosen as it is approximately in the middle of the low-dose range with regard to the dose variation \cite{IEC61526}. Furthermore, 167\,µSv has been used internally as a reference in the past for dosimetry measurements with Dosepix.

\subsection{Results for energy dependence}
\begin{figure*}[!t]
\centering
\subfloat[]{\includegraphics[width=3.in]{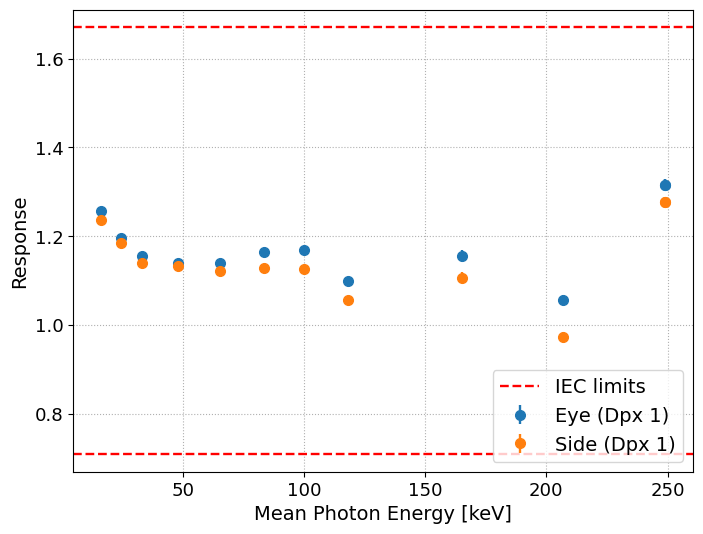}%
\label{fig:position_energy122}}
\hfil
\subfloat[]{\includegraphics[width=3.in]{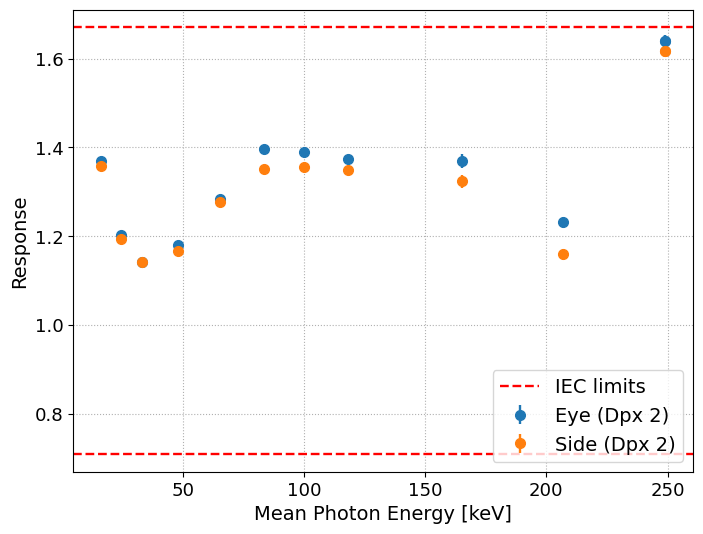}%
\label{fig:position_energy100}}
\caption{Comparison of $H_\text{p}(3)$ measured on the eye and at the side of the head of an Alderson phantom at radiation incidence of $0^\circ$, normalized to the reference dose of 167\,µSv. (a) shows the values determined by Dpx 1, (b) shows the values determined by Dpx 2. The red dashed lines mark the limits as stated in IEC 61526:2024 \cite{IEC61526}. In both figures, the relative statistical uncertainties are below 1\,\% and therefore mostly not visible.}
\label{fig:position_energy}
\end{figure*}

Fig.~\ref{fig:position_energy} shows the resulting response (measured $H_\text{p}(3)$ by irradiated reference $H_\text{p}(3)$) depending on the mean photon energy of the respective radiation qualities under an angle of incidence of $0^\circ$. 
As shown in \cite{FloPaper}, the energy dependent response differs slightly between individual Dosepix detectors and between read-out electronics. As this is not taken into account for the measurements presented in this work, the response is different compared to \cite{FloPaper}. For future investigations, those effects should be considered by adjusting the dose conversion factors accordingly for each Dosepix using a certain method that will be established in future research.\\
At low mean energies up to 48\,keV (N-60), there are only minor deviations between the values measured at the different positions for both detectors. However, the differences in response  increase as the mean photon energy increases, with the detector at the side measuring lower values than the detector on the eye. This is presumably caused by the scattering behavior of the setup, so that the discrepancy between the scattered radiation measured at the side of the Alderson phantom and on its eye increases with the mean photon energy of the radiation quality. Overall, there is a maximum difference of $(8.0\,\pm\,1.5)\,\%$ for Dpx 1 and $(5.8\,\pm\, 1.4)\,\%$ for Dpx 2. These differences are small in comparison to the maximum energy dependent variation of $(19.7\,\,\pm\,1.2)\,\%$ between the $H_\text{p}(3)$ values of N-250 and N-300 measured by Dpx 1 and $(26.7\,\pm\,0.7)\,\%$ between the $H_\text{p}(3)$ values of N-40 and N-300 measured by Dpx 2. Given the stated differences, this suggests that the intended measuring position of the active eye lens dosemeter on the side of the user's head does not significantly impair the accuracy of the dosemeter.

\subsection{Results for angular dependence}
For the examination of different irradiation angles, Dpx 1 was placed on the side of the phantom and Dpx 2 on the eye. Fig.~\ref{fig:position_energy} shows significant deviations between the two detectors at an incident angle of $0^\circ$. In order to obtain reliable results, energy dependent factors were determined from the data at $0^\circ$ by using the ratio of $H_\text{p}(3)$ measured by Dpx 1 compared to Dpx 2, both measuring on the eye. These factors were applied to the results of Dpx 2 for angled irradiation. 

\begin{figure*}[!t]
\centering
\subfloat[]{\includegraphics[width=3.in]{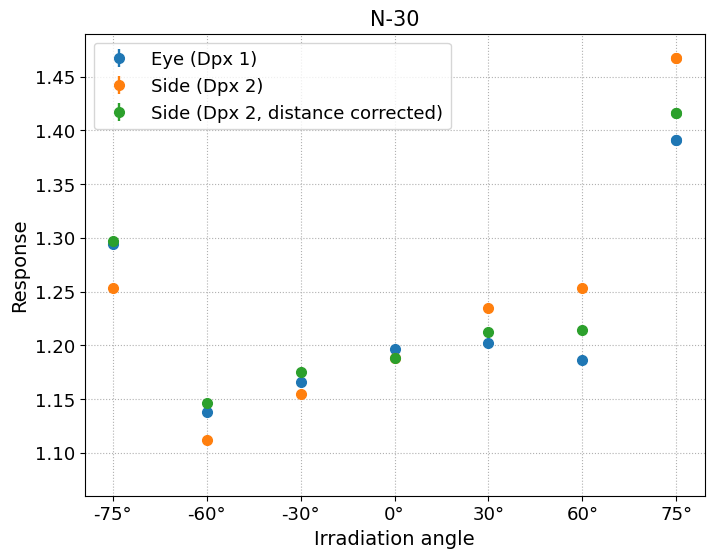}%
\label{fig:position_angle30}}
\hfil
\subfloat[]{\includegraphics[width=3.in]{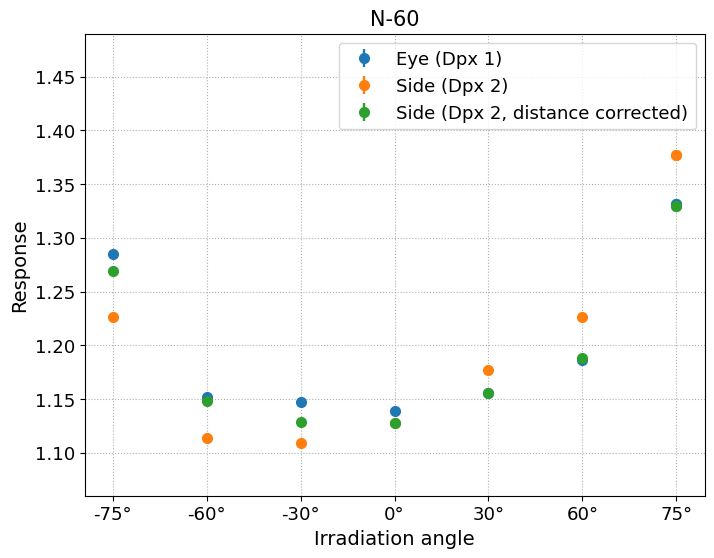}%
\label{fig:position_angle60}}
\hfil
\subfloat[]{\includegraphics[width=3.in]{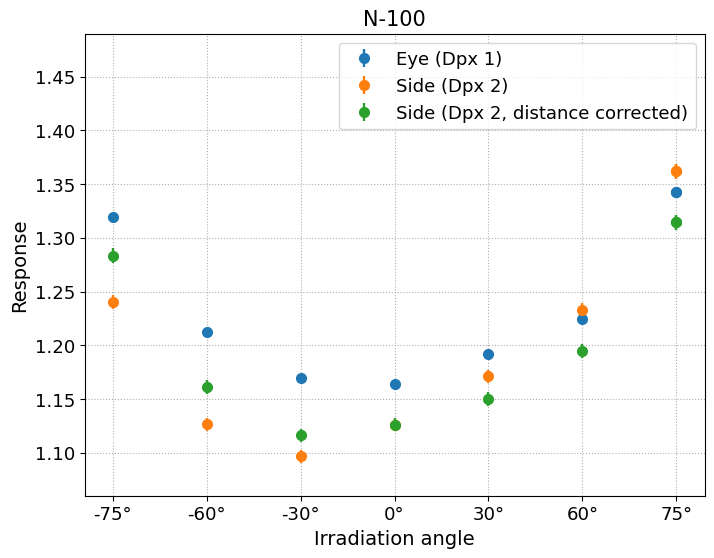}%
\label{fig:position_angle100}}
\hfil
\subfloat[]{\includegraphics[width=3.in]{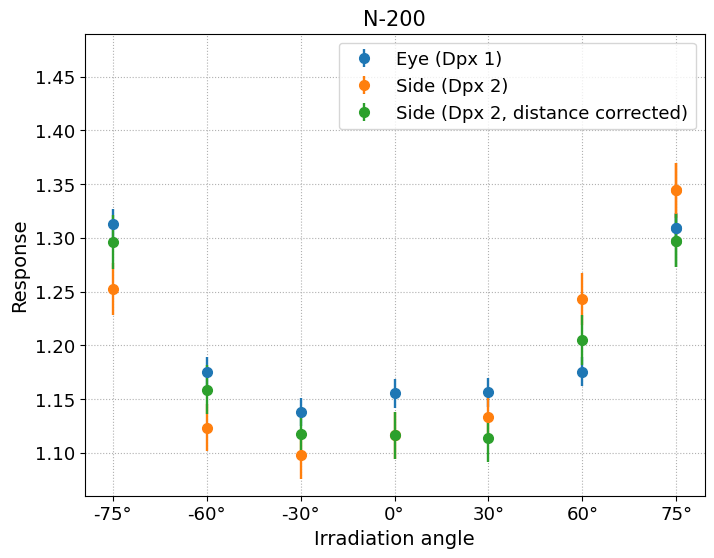}%
\label{fig:position_angle200}}
\caption{Comparison of the $H_\text{p}(3)$ response measured on the eye and the side of the head of an Alderson phantom in dependence of the irradiation angle for (a) N-30, (b) N-60, (c) N-100 and (d) N-200. The values were normalized to the reference dose of 167\,µSv. The results of Dpx 2 have been modified to account for differences between the two detectors. Additionally, the results of Dpx 2 after correcting the effect of different distances to the beam source are shown. Except for (d) due the reduced sensor sensitivity and the therefore reduced count rates, the relative statistical uncertainties are below $0.6\,\%$ for N-30, N-60 and N-100 and therefore mostly not visible. For N-200, the relative statistical uncertainties are below $2.0\,\%$.}
\label{fig:position_angle}
\end{figure*}

Fig.~\ref{fig:position_angle} shows the results for the measurements at different angles of incidence. The response values measured with the detector on the eye show an increase compared to the value measured at 0$^\circ$ for both positive and negative angles. Since the detectors are not shadowed by the phantom even at 75$^\circ$, this results from the increased scattering of the radiation in the Alderson phantom. This effect is less prominent for the measurements using N-30 as Dosepix behaves slightly differently at lower energies with respect to the angle of incidence \cite{FloMA}. On the side of the head, the resulting response values show a similar pattern, but the increase is higher for positive angles, because as the setup is turned in a positive direction, the detector on the side is closer to the beam source, i.e. closer to the X-ray tube, than the detector on the eye. Using the inverse squared relation between the expected dose and the distance to the radiation source, the difference in measured $H_\text{p}(3)$ between the detector on the side and the detector on the eye is calculated for every angle. Fig.~\ref{fig:position_angle} shows the resulting values after correcting the results of Dpx 2 with respect to these positional differences. For irradiations with N-30 and N-60, the correction brings the values measured at the side (Dpx 2) closer to the values measured at the eye (Dpx 1). However, there are larger discrepancies for N-100 and N-200. As before, this can be explained by the scattering behavior of the phantom leading to an increase of scattered radiation with higher mean photon energies. Overall, the main cause for the differences between measuring at the side and measuring at the eye seems to be the distance to the beam source for lower mean photon energies, and scattering for higher mean photon energies.

The differences between the results measured on the side and on the eye without the correction regarding the distance are between $(-7.1\,\pm 0.6)\,\%$ (N-100 at -60$^\circ$) and $(+5.70\,\pm 0.17)\,\%$ (N-30 at 60$^\circ$). This confirms the above conclusion that the measurement position on the side of the head, as intended for the active eye lens dosemeter, estimates the dose to the lens of the eye sufficiently well.

\section{Lead Glass Measurements}
\subsection{Lead Glass Pieces}
\begin{figure}
\centering
\includegraphics[width=2.7in]{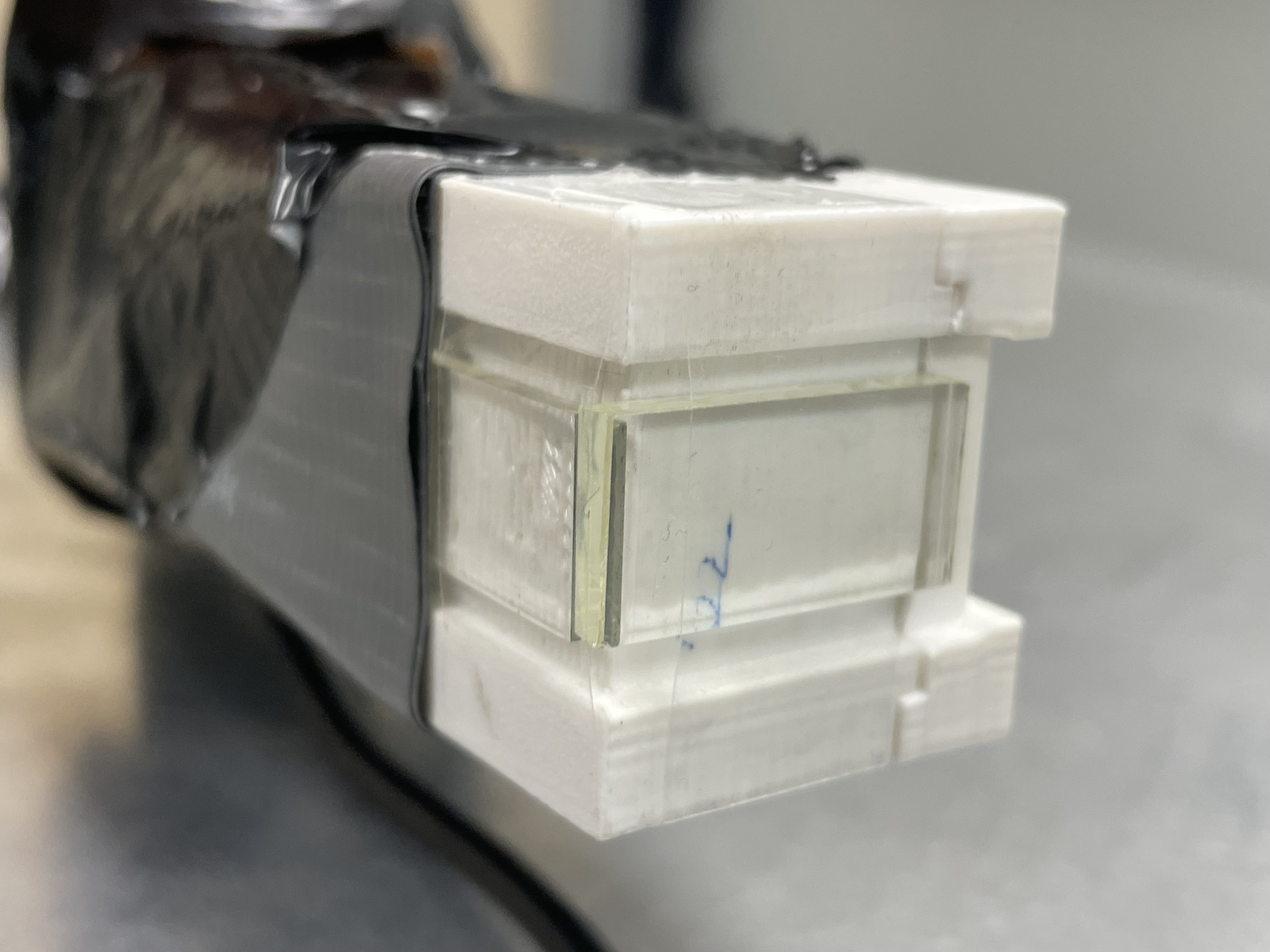}
\caption{Prototype of the active eye lens dosemeter with the frontal and side lead glass piece attached to it.}
\label{fig:ProtoLeadGlass}
\end{figure}

The application of pieces of lead glass to the eye lens dosemeter prototype in order to mimic the effect of radiation safety glasses has been previously investigated in \cite{TomBA}. An image of the eye lens dosemeter prototype with the lead glass pieces is shown in Fig.~\ref{fig:ProtoLeadGlass}. The size of the pieces was determined via intercept theorem \cite{IntTheo} by comparing the relation between the lead glass piece and Dosepix with the relation between the safety glasses and the eye lens. All values refer to the MAVIG BR126 protective glasses \cite{Mavig126}. Such radiation safety glasses usually consist of frontal lenses made of lead glass as well as side protection. This is realized by two different pieces attached to the prototype. The frontal piece has a size of $22.5 \, \text{mm} \times 11.4 \, \text{mm}$ and is affixed to the front wall of the casing, shielding the Dosepix detector directly. The second piece is smaller ($17.0 \, \text{mm} \times 11.4 \, \text{mm}$) and is attached to the side of the dosemeter prototype facing away from the user's head. Two different versions of the lead glass pieces were examined in terms of thickness. The behavior of a frontal piece of thickness of 2.5\,mm and a side piece of thickness 1.9\,mm was already examined in \cite{TomBA}. However, in this work, a frontal piece with a thickness of 2.72\,mm was used, as this reflects the MAVIG BR126 eye wear model more accurately. The thickness of the side piece remains at 1.9\,mm.

\subsection{Methods}
\begin{figure}
\centering
\includegraphics[width=2.1in]{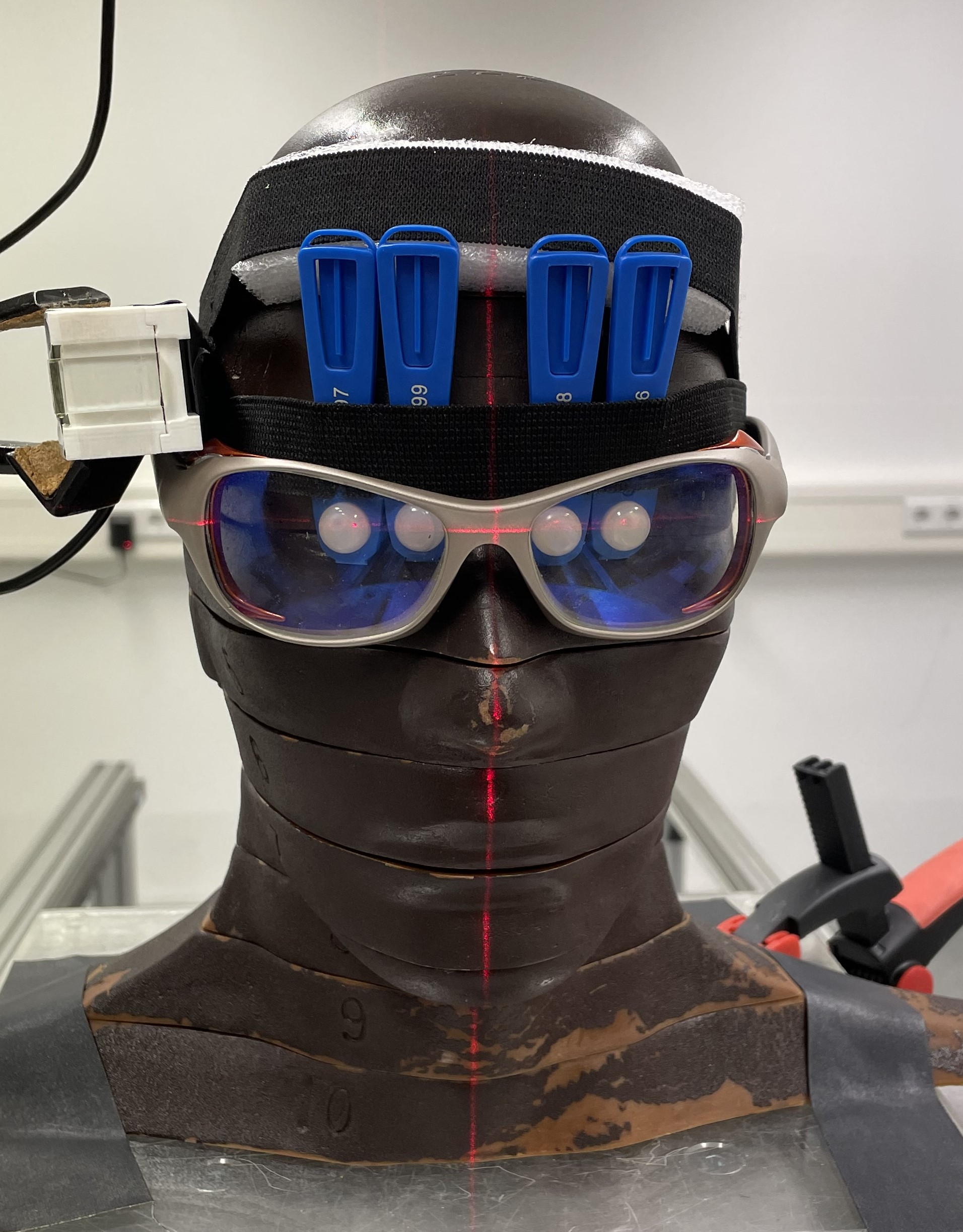}
\caption{Setup of the measurements performed at PTB. The eye lens dosemeter prototype with lead glass pieces is fixed on the side of the Alderson phantom's head. Four TLDs are placed on the eyes of the phantom behind MAVIG BR126 protective glasses. The red laser cross indicates the reference point \cite{ISO29661}.}
\label{fig:setup}
\end{figure}

To examine whether the lead glass pieces reproduce the shielding behavior of radiation safety glasses sufficiently well, several measurements were performed at the facilities of Physikalisch-Technische Bundesanstalt (PTB). A photograph of the setup is shown in Fig.~\ref{fig:setup}. The eye lens dosemeter prototype (containing Dpx 1) with lead glass pieces was placed on the side of the head of an Alderson phantom \cite{PhantomPaper} to account for the scattering behavior of a human head. For a comparison with the $H_\text{p}(3)$ behind radiation safety glasses, two EYE-D dosemeters \cite{EyeD} were placed on each eye of the Alderson phantom behind the MAVIG BR126 protective glasses. EYE-D dosemeters are passive thermoluminescent dosemeters and are hereafter referred to as TLDs. For the analysis, only the outer TLDs were taken into account as they reflect the position of the eye lens better than the two inner ones. The inner TLDs provide substitutional data in case of a faulty measurement results from the outer TLDs. The eye lens dosemeter prototype was positioned in such a manner that its Dosepix detector was in the same plane as the TLDs.

The setup was irradiated with qualities of the N-series from N-15 to N-300 \cite{ISO40371}. For N-30, N-60 and N-200, the irradiation angle was changed from 0$^\circ$ to $\pm30^\circ$, $\pm 60^\circ$ and $\pm 75^\circ$. Negative angles indicate a clockwise rotation and positive angles a counterclockwise rotation. Therefore, the eye lens dosemeter prototype in this setup is further away from the beam source at negative angles and closer to the beam source at positive angles compared to the TLDs. The differences in distances were not corrected using the inverse squared relation as before, since this corresponds to a more realistic comparison between the dose measured at the eye and at the eye lens dosemeter prototype. The setup was rotated around a reference point according to ISO 29661 \cite{ISO29661}, which was set to the nasal root of the phantom, as indicated by the red laser cross in Fig.~\ref{fig:setup}. 
A value for $H_\text{p}(3)$ of 1\,mSv was chosen for the radiation qualities N-15 to N-40 and 0.1\,mSv for N-60 to N-300. These reference doses were chosen so that the response of the TLDs is well above their detection limit. The measurement time was 100\,s for all measurements, except for N-15 with 200\,s. Since the eye lens dosemeter prototype and the TLDs were shielded by lead glass, lower measured $H_\text{p}(3)$ values than the reference dose were expected. A comparative measurement was performed without the radiation safety glasses and lead glass pieces with a reference dose of 1\,mSv.

In the following, the terms ``left'' and ``right'' refer to their medical definition, i.e. the right or left eye is always understood from the point of view of the phantom.

\subsection{Energy deposition spectra}
\begin{figure*}[!t]
\centering
\subfloat[]{\includegraphics[width=3.in]{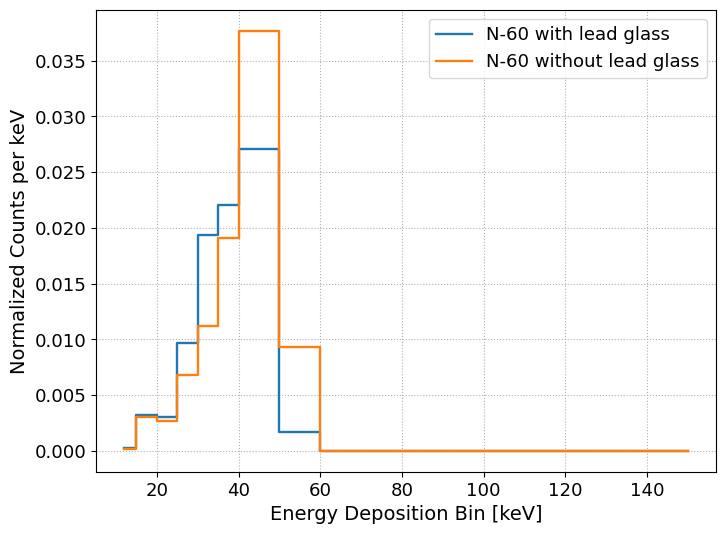}%
\label{fig:spec_N60}}
\hfil
\subfloat[]{\includegraphics[width=3.in]{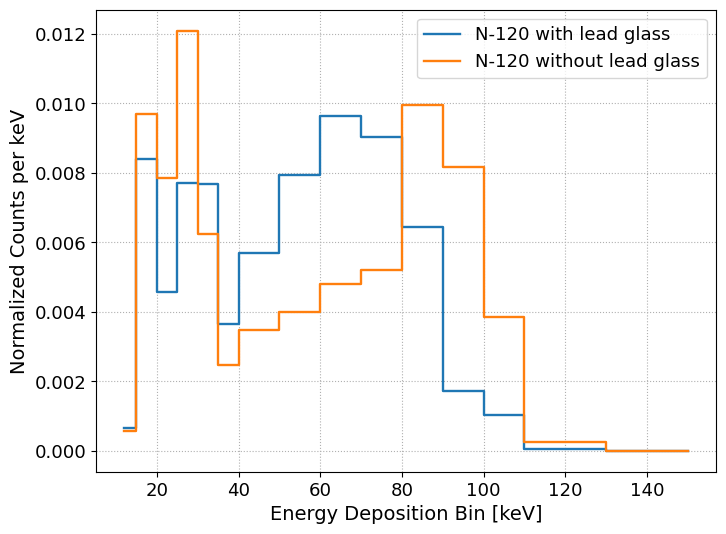}%
\label{fig:spec_N120}}
\caption{Energy deposition spectra measured by Dosepix with and without lead glass pieces for an irradiation with (a) N-60 and (b) N-120. The spectra are normalized with respect to the total number of registered events and the energy bin width.}
\label{fig:spec}
\end{figure*}

By placing lead glass in front of Dosepix, the measured energy deposition spectrum is changed. This is shown in Fig.~\ref{fig:spec} for two exemplary radiation qualities with and without lead glass pieces. The graphs show the energy deposition spectra measured by Dosepix, normalized to the total number of registered events and the width of the corresponding energy bin. Both spectra show a shift towards lower energies when measuring with lead glass. A larger relative shift is observed for N-120 than for N-60.

Theoretically, one would expect the low-energy parts of the spectrum to be filtered out by the lead glass. Since the lead glass mainly absorbs the radiation of the direct beam, the relative proportion of scattered radiation in the normalized spectrum increases. This scattered radiation consists of photons with lower energies than those in the direct beam, which leads to higher count numbers in the low energy bins. Due to normalization, the deposited spectrum appears to shift to lower energies.

\subsection{Results for energy dependence}
\begin{figure*}[!t]
\centering
\subfloat[]{\includegraphics[width=3.in]{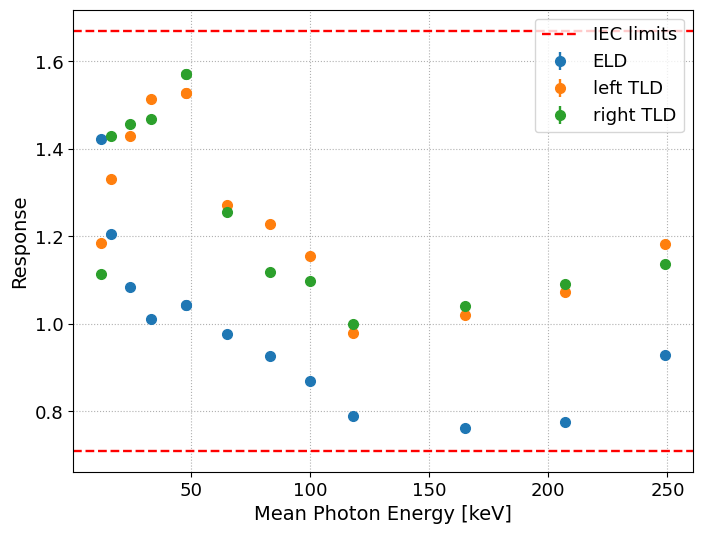}%
\label{fig:energy_blei}}
\hfil
\subfloat[]{\includegraphics[width=3.in]{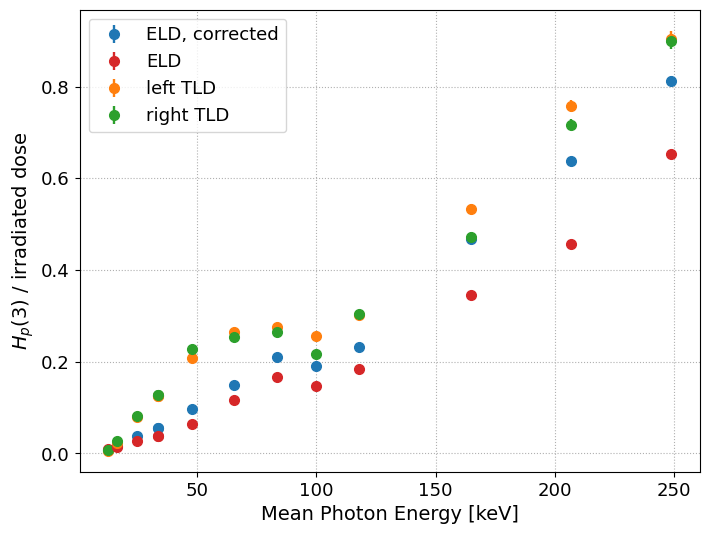}%
\label{fig:energy_keinblei}}
\caption{$H_\text{p}(3)$ response of the eye lens dosemeter prototype (ELD) and the TLDs on the left and right eye of the phantom with respect to the mean energy of the irradiated photons. (a) shows the results with lead glass pieces and radiation safety glasses. To take the detector differences between eye lens dosemeter prototype and TLDs into account, correction factors were applied to the data of the eye lens dosemeter prototype. (b) shows the results of the measurements without lead glass pieces and radiation safety glasses. The red dashed lines mark the limits as stated in IEC 61526:2024~\cite{IEC61526}. In both figures, the relative uncertainties are below 3.0\,\% and therefore not visible.}
\label{fig:energy}
\end{figure*}

A comparison of the normalized $H_\text{p}(3)$ values measured by the eye lens dosemeter prototype and the TLDs depending on the mean photon energy of the irradiated photons is shown in Fig.~\ref{fig:energy}. Fig.~\ref{fig:energy_blei} shows the results of the measurements with lead glass pieces and radiation safety glasses and Fig.~\ref{fig:energy_keinblei} the response values without the lead glass. All irradiations were performed at an angle of 0$^\circ$.

To compare the eye lens dosemeter prototype and the TLDs, measurements were performed without any lead glass shielding (see Fig.~\ref{fig:energy_keinblei}). Both detector types show a similar trend with respect to the mean photon energies. Nevertheless, it is evident that the response values of the eye lens dosemeter prototype mostly fall significantly below the response values of the TLDs. A possible explanation is given by the different energy dependencies of the two dosemeter types as well as their different positions on the Alderson phantom. To incorporate these differences in the results of the measurements with lead glass, correction factors for each energy are defined as the ratio of the $H_\text{p}(3)$ values measured by the eye lens dosemeter prototype to the average of the left and right TLDs.

Fig.~\ref{fig:energy_blei} shows the normalized $H_\text{p}(3)$ values resulting from the measurements with lead glass pieces and radiation safety glasses, with the response reconstructed directly from the measurement data of the eye lens dosemeter prototype as well as the corresponding corrected values. The latter were determined by applying correction factors calculated from the data measured without lead glass shielding, as explained above. This compensates for detector-related differences as well as different measurement positions of TLDs and eye lens dosemeter prototype. This allows the focus to be put on the influence of the lead glass. \\
The normalized $H_\text{p}(3)$ values resulting from the measurements with lead glass pieces and radiation safety glasses in Fig.~\ref{fig:energy_blei} show an overall increase with the mean photon energy for both the eye lens dosemeter prototype and the TLDs. This is explained by the lead glass becoming more transmissive to radiation for increasing energies. For both detector types, there is a local maximum at 83\,keV, i.e. for the irradiation with N-100. This is possibly explained by the K-edge of lead at 88\,keV \cite{blei}. The mean photon energy of N-100 is just below this K-edge. The absorption at this energy is therefore less than at the next higher energetic irradiation quality N-120, which is with its mean energy of 100\,keV just above the K-edge at 88\,keV and therefore undergoes an increased absorption. Therefore, the measured dose at N-100 is higher than at N-120, possibly resulting in this local maximum. \\
After applying the correction factors, the response values of the eye lens dosemeter prototype are closer to the values of the TLDs, as shown in Fig.~\ref{fig:energy_blei}.
Even after the correction, deviations up to $(57.5\,\pm 1.0)\,\%$ between the results of the two detector types remain. There are several possible explanations for this. Since the surface of the lead glass pieces is smaller than that of the radiation safety glasses, the angular range in which the radiation is influenced by the lead glass is smaller. This leads to higher $H_\text{p}(3)$ values measured by the TLDs which is consistent with the observations in Fig.~\ref{fig:energy_blei}.
Furthermore, the deposited energy spectrum of the irradiated quality changes when passing through lead glass as shown before in Fig.~\ref{fig:spec}. The mean photon energy is therefore not identical when lead glass is present or not. This energy change results in small errors since the correction factors are energy dependent. However, this influence can be assumed to be negligibly small.
Additionally, it is possible that the lead glass pieces might be of a slightly different thickness or material composition compared to the MAVIG BR126 glasses. Using the intercept theorem as a method to determine the size of the pieces might be insufficient. It is advisable to test the same methodology with other eye wear models and investigate whether similar behaviors occur.

\subsection{Results for angular dependence}
\begin{figure*}[!t]
\centering
\subfloat[]{\includegraphics[width=2.3in]{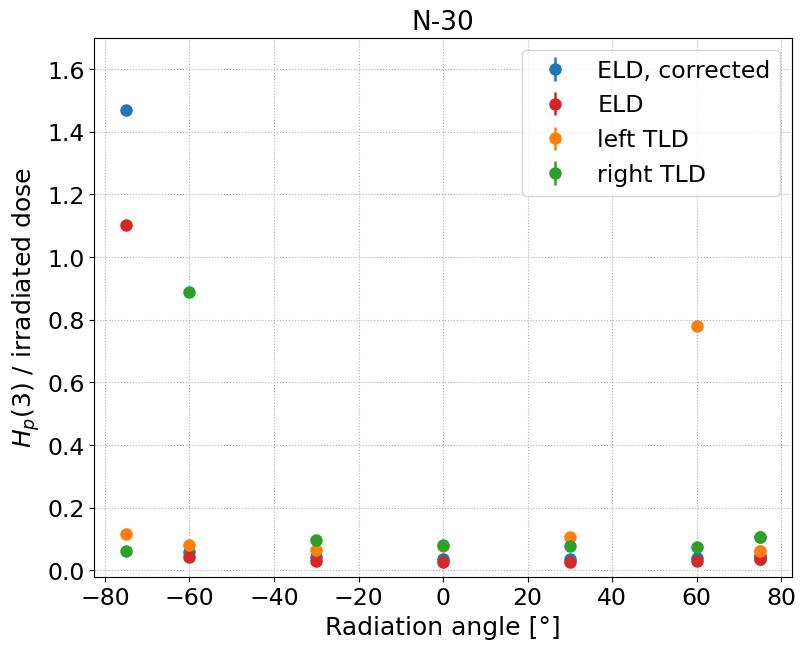}%
\label{fig:angle_N30}}
\hfil
\subfloat[]{\includegraphics[width=2.3in]{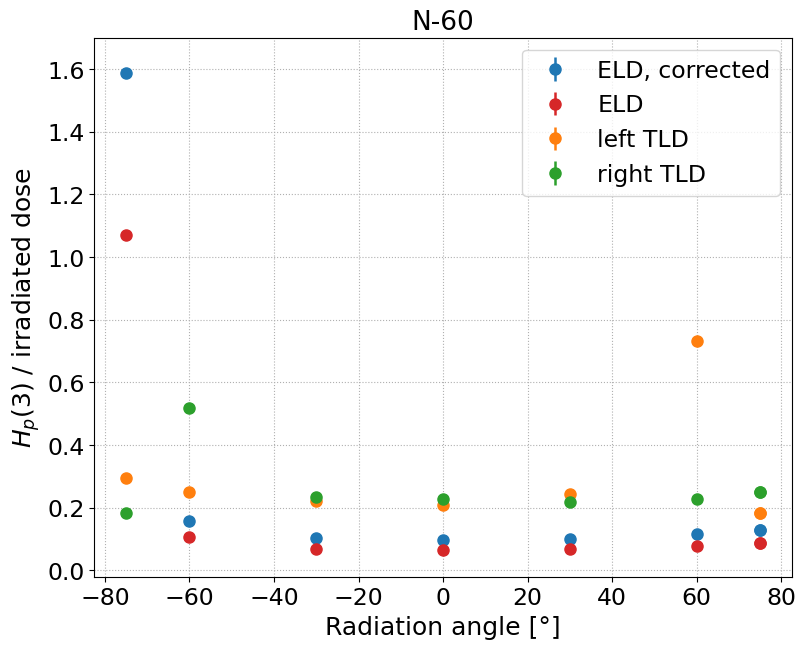}%
\label{fig:angle_N60}}
\hfil
\subfloat[]{\includegraphics[width=2.3in]{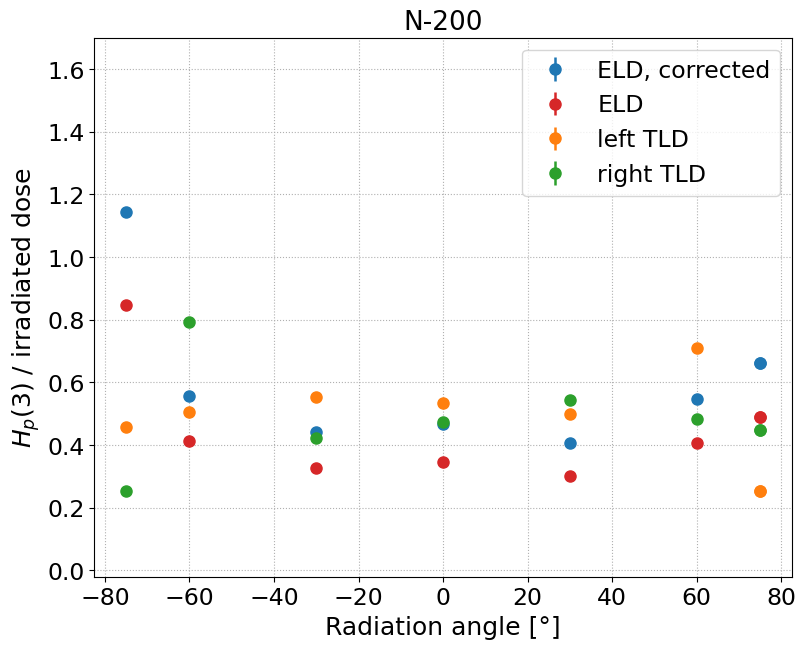}%
\label{fig:angle_N200}}
\caption{$H_\text{p}(3)$ response of the eye lens dosemeter prototype (ELD) and the TLDs on the left and right eye of the phantom with respect to the irradiation angle for (a) N-30, (b) N-60 and (c) N-200. The values are normalized to the reference dose to investigate the changes in absorption behavior under different irradiation angles and for comparability reasons. To take the detector differences between eye lens dosemeter prototype and TLDs into account, correction factors were applied to the data of the eye lens dosemeter prototype, as explained above. For all measurements, lead glass pieces and radiation safety glasses are present. In all figures, the relative uncertainties are below 2.3\,\% and therefore not visible.}
\label{fig:angle}
\end{figure*}

The measurements with lead glass pieces and radiation safety glasses were repeated for N-30, N-60 and N-200 at irradiation angles of 0$^\circ$, $\pm 30^\circ$, $\pm 60^\circ$ and $\pm 75^\circ$ to examine the change in response under different angles. The resulting $H_\text{p}(3)$ values of the eye lens dosemeter prototype and the left and right TLDs are shown in Fig.~\ref{fig:angle}, where the measured $H_\text{p}(3)$ values were normalized to the reference dose. In addition to the results measured by the eye lens dosemeter prototype (ELD), the correction factors mentioned above were applied to compensate the differences in terms of detector type and measurement position. However, it should be noted that these factors were determined only for an angle of irradiation of $0^\circ$ and therefore do not take into account the different behavior of the TLDs and the eye lens dosemeter prototype for irradiations with photons at different angles of incidence.

Again, there is an overall increase in $H_\text{p}(3)$ with the energy of the irradiated photons. However, there are some outliers. For all three qualities, the $H_\text{p}(3)$ values at $-75^\circ$ measured by the eye lens dosemeter prototype noticeably soar. This suggests that the beam bypasses the frontal lead glass piece and hits the Dosepix detector directly. Also, the uncorrected $H_\text{p}(3)$ values at $-75^\circ$ closely match the result of the unshielded measurement at that angle, further supporting this assumption.
Furthermore, the TLDs show a sharp increase at $60^\circ$ for the left TLD and at $-60^\circ$ for the right TLD, which is especially prominent for N-30 (Fig.~\ref{fig:angle_N30}) and N-60 (Fig.~\ref{fig:angle_N60}). A similar observation was made in the first examinations of the lead glass pieces \cite{TomBA}. The radiation safety glasses do not fit perfectly on the nose of the Alderson phantom but leave a small gap. Therefore, the radiation can bypass the glasses at the nasal bridge at these angles. The symmetry of the left and right TLDs further supports this assumption. 

\subsection{Material comparison}
To directly compare the material of the lead glass pieces and the MAVIG BR126 radiation safety glasses, complementary experiments were conducted at Erlangen Centre for Astroparticle Physics (ECAP). The eye lens dosemeter prototype was placed at 2\,m distance to the radiation source. The measurements were performed by placing either the radiation safety glasses (MAVIG BR126), lead glass pieces of two different thicknesses (2.5\,mm and 2.72\,mm) or a large, asymmetrical lead glass plate in front of the eye lens dosemeter prototype and irradiating it with radiation qualities of the N-series with a measurement time of 210\,s. The measurements were repeated without any shielding to obtain a reference dose.
The lead glass plate also had a thickness of 2.5\,mm with a width of $20\,\text{cm}$ and height between $7.5\,\text{cm}$ and $12 \, \text{cm}$. All materials were placed as close as possible to the eye lens dosemeter prototype. The measured dose quantity in this context is the directional dose equivalent at 3\,mm depth $H'(3)$ as no phantom was included in the setup \cite{PTBDos23}.

\begin{figure*}[!t]
\centering
\subfloat[]{\includegraphics[width=3.in]{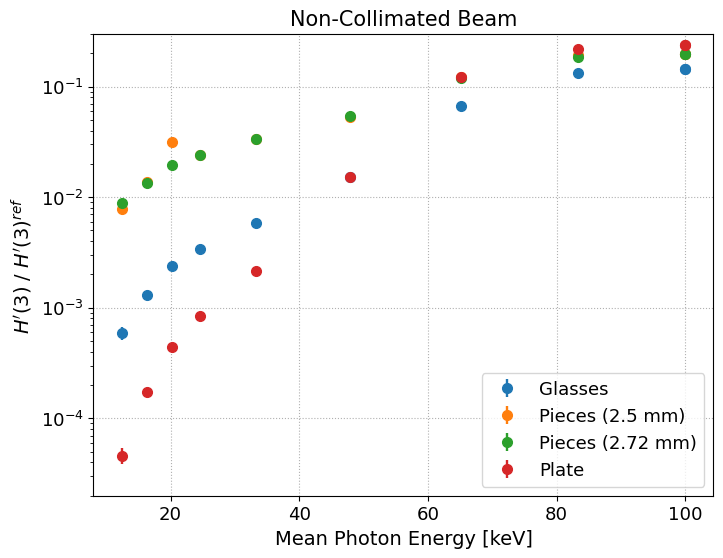}%
\label{fig:material_notcol}}
\hfil
\subfloat[]{\includegraphics[width=3.in]{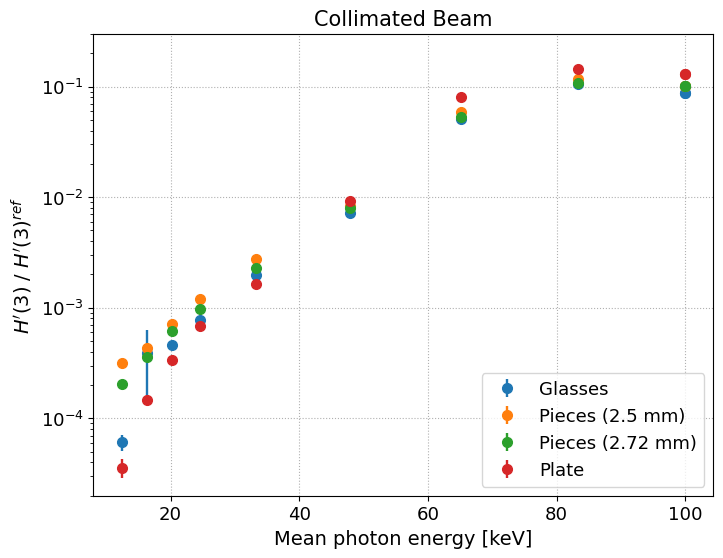}%
\label{fig:material_col}}
\caption{$H'(3)$ determined by the eye lens dosemeter prototype, normalized to the reference dose measured by the prototype without shielding. Radiation safety glasses, lead glass pieces in two different thicknesses and a lead glass plate were used as shielding material in front of the dosemeter prototype. The results are shown for both (a) a non-collimated and (b) a collimated beam. The uncertainties in both figures are small and therefore mostly not visible.}
\label{fig:material}
\end{figure*}

Fig.~\ref{fig:material} shows the results for irradiations with a collimated and a non-collimated beam. The non-collimated beam is used since this represents the realistic conditions in the medical field. For the non-collimated irradiation, the whole setup is illuminated homogeneously. The collimated beam has a diameter of 1\,cm. The measured $H'(3)$ values were normalized to the dose measured without any lead glass shielding. In the case of a non-collimated beam (Fig.~\ref{fig:material_notcol}), there are significant deviations between the measurements with radiation safety glasses, lead glass pieces and lead glass plate. This is related to the area covered by the material and the resulting influence on the scattering contribution. Since the lead glass pieces only partially shield the eye lens dosemeter prototype, more scattered radiation hits the detector from the side. On the other hand, the lead glass plate covers a large area around the dosemeter prototype resulting in less scattering and lower $H'(3)$ values. This effect is most pronounced at low mean photon energies as there, the direct beam is almost totally absorbed and, consequently, the detector's response is dominated by the scattered radiation. At higher energies, the lead glass becomes more transmissive and the differences between the materials decrease.

In order to examine whether there are differences between the materials, the scattering effect must be minimized. This is done by collimating the beam. Looking at the results in Fig.~\ref{fig:material_col}, there are only minor differences between the materials. The lead glass pieces with a thickness of 2.5\,mm tend to measure higher dose values, confirming the conclusion of \cite{TomBA} that the thickness had to be increased. The lead glass with the higher thickness of 2.72\,mm does indeed provide results that are more similar to the results of the radiation safety glasses. Despite collimating the beam, the previously observed scattering effect (Fig.~\ref{fig:material_notcol}) is still observable, especially at very low mean photon energies below 20\,keV. Not taking into account the measurement with N-15, the maximum deviations with respect to the radiation safety glasses are $(70\, \pm 13)\,\%$ (lead glass pieces 2.5\,mm), $(47\, \pm 16)\,\%$ (lead glass pieces 2.72\,mm) and $(62\, \pm 23)\,\%$ (lead glass plate). This shows that the 2.72\,mm thick lead glass pieces, as used in the previous measurements at PTB, best approximate the absorption behavior of the radiation safety glasses, but still show significant differences. The deviations between the results of the TLDs and the eye lens dosemeter prototype observed in Fig.~\ref{fig:energy_blei} could therefore originate from incorrect dimensions, thickness or different material compositions of the lead glass pieces compared to the radiation safety glasses. 

\section{Summary and Conclusion}
It has been shown in \cite{FloPaper} that the first prototype of the active eye lens dosemeter yields results within the IEC criteria in measurements of $H_\text{p}(3)$. Those measurements have been performed in compliance with IEC 61526:2024 \cite{IEC61526} in front of a cylinder phantom. The results led to interest in whether the dosemeter would also perform well when attached to the side of a phantom, similar to how it is intended to be worn by a final user. The results of this work demonstrate that there is no significant influence on the estimated dose when the measurement is performed at the side of a phantom compared to a position at the center front, for mean photon energies ranging from 12.4\,keV to 249\,keV as well as for angles between $-75^\circ$ and $75^\circ$. However, it has to be pointed out, that these findings were obtained in rather homogeneous radiation fields since the radiation source (X-ray tube) was located at a distance of 2.5\,m from the irradiation setup. Thus, corresponding investigations should be repeated in more realistic radiology situations, for instance approximately 50\,cm away from a stray body, i.e. a patient, in real life.

Since medical staff usually wear radiation safety glasses when working with ionizing radiation, an attempt was presented to imitate the absorption behavior of such glasses. Small lead glass pieces were attached to the front and the side of the eye lens dosemeter prototype. Additionally, TLDs were installed behind radiation safety glasses for comparison. When irradiated with photons of different angles of incidence, the results of the TLDs showed that the radiation is able to pass through a gap between the nose and the radiation safety glasses at an angle of $\pm 60^\circ$. For an irradiation angle of $-75^\circ$, the results of the eye lens dosemeter prototype showed that the radiation bypasses the front lead glass piece and hits the Dosepix detector unshielded. There was a consistent offset between the measured $H_\text{p}(3)$ of the TLDs and the eye lens dosemeter prototype, with the prototype measuring lower values. There are several possible explanations for this effect. First, the correction factors used to eliminate the detector differences are energy dependent due to the absorption of the lead glasses affecting the incoming spectra. This effect was considered to be small and was therefore neglected in the analysis. Second, but more significant, is the influence exerted by the lead glass pieces themselves. Comparative measurements were performed to examine the differences between radiation safety glasses and lead glass pieces. The results revealed a strong influence from scattered radiation related to the area shielded by the respective material. The effect was significantly smaller, but still visible with a collimated beam. It is therefore reasonable to assume that these differences between lead glass pieces and radiation safety glasses are responsible for the measured differences of TLDs and eye lens dosemeter prototype. However, while the prototype with lead glass pieces measured lower dose values than the TLDs behind radiation safety glasses, the material comparison showed the opposite behavior. There, the resulting values of the measurements with lead glass pieces were higher than those of the measurements with radiation safety glasses. Ultimately, many different setup properties are possibly influencing the final dose reconstruction. It is still possible that the lead glass pieces are made from a different material since the exact material composition of the lead glass is unknown, or that they have different thicknesses or dimensions. The intercept theorem used for determining the size of the lead glass pieces might not be sufficient. This could be examined by repeating the process with different models of radiation safety glasses. It is also possible that the TLDs measure more backscattered radiation from the phantom. Since the rear of the TLDs consists of plastic, similarly to the front, the sensitivity is approximately equal from both sides. On the other hand, it has been shown that the response of eye lens dosemeter prototype increases only by up to 20\,\% when comparing the measurement position at the side of a water cylinder phantom to the phantom being behind the dosemeter \cite{FloMA}. This indicates that the eye lens dosemeter prototype is less sensitive to scattered radiation than the TLDs.
To investigate this effect, a measurement without an Alderson phantom would be advisable. Another possibility is to exchange the EYE-D dosemeter with other passive dosemeters and compare the results with the eye lens dosemeter prototype.

Overall, the positioning of the eye lens dosemeter has proved to be of lower significance, but the reason for the differences between shielding glasses, differences between the eye lens dosemeter prototype and the TLDs as well as the question of which of the two detector types comes closest to the actual dose in the eye lens is still unclear. More research is required, but the use of lead glass pieces is a very promising approach.

\section*{Acknowledgments}
The authors declare that they have no known conflicts of interest in terms of competing financial interests or personal relationships that could have an influence or are relevant to the work reported in this article. The authors are grateful to Christian Fuhg (PTB) for his valuable support during the measurements and George Winterbottom (PTB) for the language check.

\vfill

\end{document}